\documentclass{article}
\usepackage{amssymb}
\usepackage{amsmath}

\input{tcilatex}
\begin{document}

\begin{center}

{\LARGE Properties of the zeros of generalized basic hypergeometric
polynomials}

\bigskip

$^{\ast }$\textbf{Oksana Bihun}$^{1}$ and $^{+\lozenge }$\textbf{Francesco
Calogero}$^{2}\bigskip $

$^{\ast }$Department of Mathematics, Concordia College\\[0pt]
901 8th Str. S, Moorhead, MN 56562, USA, +1-218-299-4396 \smallskip

$^{+}$Physics Department, University of Rome \textquotedblleft La Sapienza"\\%
[0pt]
p. Aldo Moro, I-00185 ROMA, Italy, +39-06-4991-4372 \smallskip

$^{\lozenge }$Istituto Nazionale di Fisica Nucleare, Sezione di Roma
\smallskip

$^{1}$obihun@cord.edu

$^{2}$francesco.calogero@roma1.infn.it, francesco.calogero@uniroma1.it

\bigskip

\textit{Abstract}
\end{center}

We define the \textit{generalized basic hypergeometric polynomial} of degree 
$N$ as follows:%
\begin{eqnarray*}
&&P_{N}\left( \alpha _{1},...,\alpha _{r};\beta _{1},...,\beta
_{s};q;z\right) =\sum_{m=0}^{N}\left[ \frac{\left( q^{-N};q\right)
_{m}\left( \alpha _{1};q\right) _{m}\cdot \cdot \cdot \left( \alpha
_{r};q\right) _{m}}{\left( q;q\right) _{m}~\left( \beta _{1};q\right)
_{m}\cdot \cdot \cdot \left( \beta _{s};q\right) _{m}}\left[ \left(
-1\right) ^{m}~q^{m\left( m-1\right) /2}\right] ^{s-r}~z^{m}\right] \\
&=&_{r+1}\phi _{s}\left( q^{-N},\alpha _{1,}...,\alpha _{r};\beta
_{1},...,\beta _{s};q;z\right) ~.
\end{eqnarray*}%
Here $N$ is an arbitrary \textit{positive} integer, $r$ and $s$ are
arbitrary \textit{nonnegative} integers, the $r+s$ parameters $\alpha _{j}$
and $\beta _{k}$ are arbitrary (\textquotedblleft generic\textquotedblright
, possibly complex) numbers, $\left( \alpha ;q\right) _{m}$ is the $q$%
-Pochhammer symbol and $_{r+1}\phi _{s}\left( \alpha _{0},\alpha
_{1,}...,\alpha _{r};\beta _{1},...,\beta _{s};q;z\right) $ is the
generalized basic hypergeometric function. In this paper we obtain a set of $%
N$ \textit{nonlinear algebraic equations} satisfied by the $N$ zeros $\zeta
_{n}\equiv \zeta _{n}\left( \underline{\alpha },\underline{\beta }%
;q;N\right) $ of this polynomial. We moreover identify an $\left( N\times
N\right) $-matrix $\underline{M}\equiv \underline{M}\left( \underline{\alpha 
},\underline{\beta };\underline{\zeta };q;N\right) $ featuring the $N$
eigenvalues $\mu _{n}=-q^{\left( s-r\right) \left( N-n\right) }\left(
q^{-n}-1\right) ~\dprod\limits_{j=1}^{r}\left( \alpha _{j}~q^{N-n}-1\right)
\,,~~~n=1,2,...,N~.$ These $N$ eigenvalues depend only on the $r$ parameters 
$\alpha _{j}$ (besides $q$ and $N$), implying that the $\left( N\times
N\right) $-matrix $\underline{M}$ is \textit{isospectral} for variations of
the $s$ parameters $\beta _{k}$; and they clearly are \textit{rational}
numbers if $q$ and the $r$ parameters $\alpha _{j}$ are themselves \textit{%
rational} numbers: a nontrivial \textit{Diophantine} property.

\bigskip


\bigskip

\section{Introduction}

We define the \textit{generalized basic hypergeometric polynomial} of degree 
$N$ as follows: 
\begin{eqnarray}
&&P_{N}\left( \alpha _{1},...,\alpha _{r};\beta _{1},...,\beta
_{s};q;z\right) =\sum_{m=0}^{N}\left\{ \frac{\left( q^{-N};q\right)
_{m}\left( \alpha _{1};q\right) _{m}\cdot \cdot \cdot \left( \alpha
_{r};q\right) _{m}}{\left( q;q\right) _{m}~\left( \beta _{1};q\right)
_{m}\cdot \cdot \cdot \left( \beta _{s};q\right) _{m}}\left[ \left(
-1\right) ^{m}~q^{m\left( m-1\right) /2}\right] ^{s-r}~z^{m}\right\}  \notag
\\
&=&_{r+1}\phi _{s}\left( q^{-N},\alpha _{1,}...,\alpha _{r};\beta
_{1},...,\beta _{s};z\right) ~.  \label{PN}
\end{eqnarray}%
Here $N$ is an arbitrary \textit{positive} integer, $r$ and $s$ are
arbitrary \textit{nonnegative} integers, the $r+s$ parameters $\alpha _{j}$
and $\beta _{k}$ are arbitrary (\textit{generic}, possibly \textit{complex})
numbers, $\left( \gamma ;q\right) _{m}$ is the $q$-Pochhammer symbol, 
\begin{equation}
\left( \gamma ;q\right) _{0}=1~,~~~\left( \gamma ;q\right) _{m}=\left(
1-\gamma \right) ~\left( 1-\gamma q\right) ~\left( 1-\gamma q^{2}\right)
\cdot \cdot \cdot \left( 1-\gamma q^{m-1}\right) ~~~\text{for}%
~~~m=1,2,3,...~,  \label{qPoch}
\end{equation}%
and $_{r+1}\phi _{s}\left( \alpha _{0},\alpha _{1,}...,\alpha _{r};\beta
_{1},...,\beta _{s};q;z\right) $ is the generalized basic hypergeometric
function (see (\ref{GenBasicHyp}) below for its definition, of course
consistent with (\ref{PN}); and, for instance, \cite{GR1990} for its
properties). We actually prefer to work hereafter with the \textit{monic}
version of this polynomial, $p_{N}\left( \underline{\alpha };\underline{%
\beta };q;z\right) ,$ defined of course as follows: 
\begin{subequations}
\label{Monic}
\begin{equation}
p_{N}\left( \underline{\alpha };\underline{\beta };q;z\right) =\frac{\left(
q;q\right) _{N}~\left( \beta _{1};q\right) _{N}\cdot \cdot \cdot \left(
\beta _{s};q\right) _{N}}{\left( q^{-N};q\right) _{N}~\left( \alpha
_{1};q\right) _{N}\cdot \cdot \cdot \left( \alpha _{r};q\right) _{N}}\left[
\left( -1\right) ^{N}~q^{N\left( N-1\right) /2}\right] ^{r-s}~P_{N}\left( 
\underline{\alpha };\underline{\beta };q;z\right) ~,  \label{pN}
\end{equation}%
which is clearly characterized by the property 
\begin{equation}
\underset{z\rightarrow \infty }{\lim }\left[ \frac{p_{N}\left( \underline{%
\alpha };\underline{\beta };q;z\right) }{z^{N}}\right] =1~.  \label{pNmonic}
\end{equation}

In this paper we report properties satisfied by the $N$ zeros $\zeta
_{n}\equiv \zeta _{n}\left( \underline{\alpha };\underline{\beta }%
;q;N\right) $ of this polynomial $p_{N}\left( z\right) $ (or, equivalently, $%
P_{N}\left( z\right) $), namely by the $N$ roots $\zeta _{n}$ defined (up to
permutations) by the formula 
\end{subequations}
\begin{subequations}
\begin{equation}
p_{N}\left( \underline{\alpha };\underline{\beta };q;\zeta _{n}\right)
=0~,~~~n=1,2,...,N~,  \label{Zeros}
\end{equation}%
so that%
\begin{equation}
p_{N}\left( \underline{\alpha };\underline{\beta };q;z\right)
=\prod\limits_{n=1}^{N}\left[ z-\zeta _{n}\left( \underline{\alpha };%
\underline{\beta };q;N\right) \right] ~.  \label{Zerosb}
\end{equation}

\textbf{Notation 1.1}. Let us confirm that throughout this paper the
parameters $r$, $s$ and $N$ are $3$ \textit{arbitrary} nonnegative integers
(except when their values are explicitly assigned, see for instance the
subsections at the end of next section). Above (and hereafter) the
short-hand notations $\underline{\alpha }$, $\underline{\beta }$
respectively $\underline{\zeta }$ denote the \textit{unordered} sets of the $%
r$, $s$ respectively $N$ numbers $\alpha _{j}$, $\beta _{k}$ respectively $%
\zeta _{n}$. The $N$ zeros $\zeta _{n}$ are of course functions of the $%
r+s+2 $ parameters $\alpha _{j}$, $\beta _{k}$, $q\neq 1$ and $N$, i. e. $%
\underline{\zeta }\equiv \underline{\zeta }\left( \underline{\alpha };%
\underline{\beta };q;N\right) $. Note that we occasionally omit to indicate
explicitly the dependence on some, or on all, of these parameters
(including, systematically, the dependence on $r$ and $s$, the values of
which are considered fixed throughout this paper). Generally (unless
otherwise specified) the indices $n$, $m$, $\ell $ run over the positive
integers from $1$ to $N$, while the index $j$ runs from $1$ to $r$ and the
index $k$ from $1$ to $s$. Upper-case underlined letters denote $N\times N$
matrices: for instance the matrix $\underline{M}$ has the $N^{2}$ elements $%
M_{nm}$. We moreover assume the parameters $\alpha _{j}$, $\beta _{k}$, $%
q\neq 1$ to have \textit{generic} (possibly \textit{complex}) values, and
the $N$ zeros $\zeta _{n}$ to be all different among themselves, $\zeta
_{n}\neq \zeta _{m}$ for $n\neq m$, which is of course the \textit{generic}
case; on the understanding that otherwise our formulas might have to be
understood \textit{cum grano salis}, via appropriate limiting processes. $%
\square $

In this paper we firstly obtain a set of algebraic equations satisfied by
the $N$ zeros $\zeta _{n}$ of the \textit{generalized basic hypergeometric
polynomial} of order $N$, additional to the algebraic equation defining
these zeros, see (\ref{Zeros}). This result---reported below as \textbf{%
Proposition 2.1} because we were unable to find any mention of it in the
literature---does not seem to us to be particularly significant, but it is
instrumental to prove what we consider our main finding: the identification
of an $\left( N\times N\right) $-matrix $\underline{M}$ that features the $N$
eigenvalues 
\end{subequations}
\begin{equation}
\mu _{n}=-q^{\left( s-r\right) \left( N-n\right) }\left( q^{-n}-1\right)
~\dprod\limits_{j=1}^{r}\left( \alpha _{j}~q^{N-n}-1\right)
\,,~~~n=1,2,...,N~.~\square
\end{equation}%
This matrix $\underline{M}$ is explicitly defined below in terms of the $%
r+s+2$ parameters $\alpha _{j}$, $\beta _{k}$, $q\neq 1$ and $N$
characterizing the \textit{generalized basic hypergeometric polynomial }$%
p_{N}\left( z\right) $, and of its $N$ zeros $\zeta _{n}$ (which themselves
depend of course on the $r+s+2$ parameters $\alpha _{j}$, $\beta _{k}$, $%
q\neq 1$ and $N$). While the $N$ eigenvalues $\mu _{n}$ clearly depend only
on the $r+2$ parameters $\alpha _{j}$, $q\neq 1$\textbf{\ }and $N$, implying
that the $\left( N\times N\right) $-matrix $\underline{M}$ is \textit{%
isospectral} for variations of the $s$ parameters $\beta _{k}$; and \textit{%
all} these eigenvalues are clearly \textit{rational} numbers if $q$ and the $%
r$ parameters $\alpha _{j}$ are themselves \textit{rational} numbers: a
nontrivial \textit{Diophantine} property.

The findings outlined above are detailed in the following Section 2, and our
main finding is proven in Section 3. A terse Section 4 (``Outlook'')
outlines possible future developments. The definitions and some standard
properties of the generalized basic hypergeometric polynomials are reported
in the Appendix, for the convenience of the reader and also to specify our
notation; the reader is advised to glance through this Appendix before
reading the next section, and then to return to it whenever appropriate.

For a terse review of analogous results for \textit{generalized
hypergeometric polynomials} the interested reader is referred to \cite%
{BC2014}; of course the results reported in this reference \cite{BC2014} can
be obtained (up to obvious notational changes) from those reported in the
present paper by taking appropriately the $q\rightarrow 1$ limit. For
results analogous to those obtained here, but for polynomials belonging to
the Askey and $q$-Askey schemes, see \cite{BC2015a} \cite{BC2015b}; and let
us recall in this connection that the polynomials belonging to the Askey,
respectively the $q$-Askey, schemes (see for instance \cite{KS}) are also
special cases of generalized hypergeometric, respectively basic
hypergeometric, functions, with indices $r\leq 3$ and $s\leq 3,$ in which
the polynomial variable is however related to \textit{parameters} of the
hypergeometric function rather than to its \textit{argument}; hence the
results reported in \cite{BC2015a} \cite{BC2015b} are \textit{not} special
cases of those obtained and reported in \cite{BC2014} and below.

Finally, let us mention that the properties of the zeros of polynomials are
a core problem of mathematics to which, over time, an immense number of
investigations have been devoted. Nevertheless new findings in this area
continue to emerge, see, for instance \cite{BCD} \cite{CI2013} \cite{IR2013} 
\cite{CY} \cite{BCY} \cite{C} (and of course \cite{BC2014} \cite{BC2015a} 
\cite{BC2015b}).

\section{Results}

The generalized basic hypergeometric function $_{r+1}\phi _{s}\left( \alpha
_{0},\alpha _{1,}...,\alpha _{r};\beta _{1},...,\beta _{s};q;z\right) $ is
defined as follows (see for instance eq. (1.2.22) in \cite{GR1990}, or eq.
(0.4.2) in \cite{KS}): 
\begin{equation}
_{r+1}\phi _{s}\left( \alpha _{0},\alpha _{1,}...,\alpha _{r};\beta
_{1},...,\beta _{s};q;z\right) =\sum_{p=0}^{\infty }\left[ \frac{\left(
\alpha _{0};q\right) _{p}\left( \alpha _{1};q\right) _{p}\cdot \cdot \cdot
\left( \alpha _{r};q\right) _{p}}{\left( q;q\right) _{p}~\left( \beta
_{1};q\right) _{p}\cdot \cdot \cdot \left( \beta _{s};q\right) _{p}}~\left[
\left( -1\right) ^{p}~q^{p\left( p-1\right) /2}\right] ^{s-r}~z^{p}\right] ~,
\label{GenBasicHyp}
\end{equation}%
where the $q$-Pochhammer symbol $\left( \alpha ;q\right) _{p}$ is defined
above, see (\ref{qPoch}), and the rest of the notation is, we trust,
self-evident. Clearly if one of the $r+1$ parameters $\alpha _{j}$ coincides
with a negative integer power of $q$, say (without loss of generality) $%
\alpha _{0}=q^{-N}$, and all the other $r+s$ parameters $\alpha _{j}$ and $%
\beta _{k}$ have \textit{generic} (possibly \textit{complex}) values, the
series in the right-hand side of the definition (\ref{GenBasicHyp}) of the
generalized hypergeometric function terminates at $j=N$ (since clearly $%
\left( q^{-N};q\right) _{p}=0$ for $p=N+1,~N+2,...,$ see (\ref{qPoch})).
This leads to the definition (\ref{PN}) of the \textit{generalized basic
hypergeometric polynomial} $P_{N}\left( z\right) $ (of degree $N$ in $z$;
below, we rather work with the \textit{monic} version $p_{N}\left( z\right) $
of this polynomial, see (\ref{Monic})).

The first result of this paper---proven in the Appendix---is the following

\textbf{Proposition 2.1}. The $N$ zeros $\zeta _{n}\equiv \zeta _{n}\left( 
\underline{\alpha };\underline{\beta };q;N\right) $ of the generalized basic
hypergeometric (monic) polynomial $p_{N}\left( \underline{\alpha };%
\underline{\beta };q;z\right) ,$ see (\ref{PN}), (\ref{pN}) and (\ref{Zeros}%
), satisfy the following set of $N$ algebraic equations: 
\begin{eqnarray}
-\prod\limits_{m=1}^{N}\left( \zeta _{n}~q-\zeta _{m}\right)
+\sum_{k=1}^{s}\left( -q\right) ^{-k}b_{k}\left( \underline{\beta }\right) ~%
\left[ \prod\limits_{m=1}^{N}\left( \zeta _{n}~q^{k}-\zeta _{m}\right)
-\prod\limits_{m=1}^{N}\left( \zeta _{n}~q^{k+1}-\zeta _{m}\right) \right] &&
\notag \\
-\left( -1\right) ^{r-s}~\zeta _{n}~\left[ \prod\limits_{m=1}^{N}\left(
\zeta _{n}~q^{s-r}-\zeta _{m}\right) -q^{-N}\prod\limits_{m=1}^{N}\left(
\zeta _{n}~q^{s-r+1}-\zeta _{m}\right) \right] &&  \notag \\
-\left( -1\right) ^{r-s}~\zeta _{n}~\left\{ \sum_{j=1,~j\neq r-s}^{r}\left[
\left( -1\right) ^{j}a_{j}\left( \underline{\alpha }\right)
~\prod\limits_{m=1}^{N}\left( \zeta _{n}~q^{s-r+j}-\zeta _{m}\right) \right]
\right. &&  \notag \\
\left. -q^{-N}\sum_{j=1,~j\neq r-s-1}^{r}\left[ \left( -1\right)
^{j}a_{j}\left( \underline{\alpha }\right) ~~\prod\limits_{m=1}^{N}\left(
\zeta _{n}~q^{s-r+j+1}-\zeta _{m}\right) \right] \right\} =0~, &&  \notag \\
n=1,2,..,N~. &&  \label{Prop1}
\end{eqnarray}
Here the $s$ quantities $b_{k}\left( \underline{\beta }\right) ,$
respectively the $r$ quantities $a_{j}\left( \underline{\alpha }\right) ,$
are defined, in terms of the $s$ parameters $\beta _{k}$ respectively the $r$
parameters $\alpha _{j},$ by (\ref{bk}) respectively (\ref{aj}). Note that
the formula (\ref{Prop1}) becomes a bit simpler in the \textquotedblleft
balanced\textquotedblright\ case (i. e., if $r=s$). Also note that, if
instead $s>r$, some terms in the sum over the index $j$ have a vanishing
value (for $j=s-r$ or $j=s-r-1$). $\square $

Our main finding---proven in Section 3---is the following

\textbf{Proposition 2.2}. Let the $\left( N\times N\right) $-matrix $%
\underline{M}$ be defined, componentwise, as follows in terms of the $r+s+2$
parameters $\alpha _{j}$, $\beta _{k}$, $q$, $N$ characterizing the
generalized basic hypergeometric polynomial and of its $N$ zeros $\zeta
_{n}\equiv \zeta _{n}\left( \underline{\alpha };\underline{\beta }%
;q;N\right) $, see (\ref{PN}), (\ref{pN}) and (\ref{Zeros}): 
\begin{subequations}
\label{MBis}
\begin{eqnarray}
&&M_{nn}=(-1)^{s}\Bigg\{(q-1)^{2}g_{n}(1,\underline{\zeta }%
)+\sum_{k=1}^{s}b_{k}(\underline{\beta })\frac{(-1)^{k}}{q^{k}}\big[%
(q^{k+1}-1)^{2}g_{n}(k+1,\underline{\zeta })-(q^{k}-1)^{2}g_{n}(k,\underline{%
\zeta })\big]\Bigg\}  \notag \\
&&+(-1)^{r+1}\zeta _{n}\Bigg\{q^{-N}(q^{s-r+1}-1)^{2}g_{n}(s-r+1,\underline{%
\zeta })-(q^{s-r}-1)^{2}g_{n}(s-r,\underline{\zeta })  \notag \\
&&+\sum_{j=1}^{r}a_{j}(\underline{\alpha })(-1)^{j}\big[%
q^{-N}(q^{j+s+1-r}-1)^{2}g_{n}(j+s+1-r,\underline{\zeta }%
)-(q^{j+s-r}-1)^{2}g_{n}(j+s-r,\underline{\zeta })\big]\Bigg\}  \notag \\
&&+(-1)^{r}\Bigg\{q^{-N}(q^{s-r+1}-1)f_{n}(s-r+1,\underline{\zeta }%
)-(q^{s-r}-1)f_{n}(s-r,\underline{\zeta })  \notag \\
&&+\sum_{j=1}^{r}a_{j}(\underline{\alpha })(-1)^{j}\Big[%
q^{-N}(q^{j+s+1-r}-1)f_{n}(j+s+1-r,\underline{\zeta }%
)-(q^{j+s-r}-1)f_{n}(j+s-r,\underline{\zeta })\Big]\Bigg\}~,  \notag \\
&&n=1,2,...,N,  \label{MnnBis}
\end{eqnarray}%
\begin{eqnarray}
&&M_{nm}=(-1)^{s+1}\frac{\zeta _{n}}{(\zeta _{n}-\zeta _{m})^{2}}\Bigg\{%
(q-1)^{2}f_{nm}(1,\underline{\zeta })  \notag \\
&&+\sum_{k=1}^{s}b_{k}(\underline{\beta })\frac{(-1)^{k}}{q^{k}}\big[%
(q^{k+1}-1)^{2}f_{nm}(k+1,\underline{\zeta })-(q^{k}-1)^{2}f_{nm}(k,%
\underline{\zeta })\big]\Bigg\}  \notag \\
&&+(-1)^{r}\frac{\zeta _{n}^{2}}{(\zeta _{n}-\zeta _{m})^{2}}\Bigg\{%
q^{-N}(q^{s-r+1}-1)^{2}f_{nm}(s-r+1,\underline{\zeta }%
)-(q^{s-r}-1)^{2}f_{nm}(s-r,\underline{\zeta })  \notag \\
&&+\sum_{j=1}^{r}a_{j}(\underline{\alpha })(-1)^{j}\big[%
q^{-N}(q^{j+s+1-r}-1)^{2}f_{nm}(j+s+1-r,\underline{\zeta }%
)-(q^{j+s-r}-1)^{2}f_{nm}(j+s-r,\underline{\zeta })\big]\Bigg\},  \notag \\
&&n,m=1,2,...,N,~~~n\neq m~,  \label{MnmBis}
\end{eqnarray}%
where the quantities $f_{nm}(p,\underline{\zeta })$ and $g_{n}(p,\underline{%
\zeta })$ are defined by~(\ref{fmngn}). Then the $N$ eigenvalues $\mu _{n}$
of this matrix are given by the following neat formulas: 
\end{subequations}
\begin{equation}
\mu _{n}=-q^{\left( s-r\right) \left( N-n\right) }\left( q^{-n}-1\right)
~\dprod\limits_{j=1}^{r}\left( \alpha _{j}~q^{N-n}-1\right)
\,,~~~n=1,2,...,N~.~\square  \label{mun}
\end{equation}

The following corollary, which is an immediate consequence of \textbf{%
Proposition 2.2}, yields, via the definition (\ref{MBis}) of the $\left(
N\times N\right) $-matrix $\underline{M}$ and the expression (\ref{mun}) of
its $N$ eigenvalues $\mu _{n}$, a number of algebraic formulas satisfied by
the $N$ zeros $\zeta _{n}$ of the generalized basic hypergeometric
polynomial of order $N,$ see (\ref{PN}) or (\ref{pN}) and (\ref{Zeros}).

\textbf{Corollary 2.1}. 
\begin{subequations}
\begin{equation}
\text{Trace}\left[ \left( \underline{M}\right) ^{p}\right] =\sum_{n=1}^{N}%
\left[ \left( \mu _{n}\right) ^{p}\right] ~,~~~p=1,2,3,...~,
\end{equation}%
\begin{equation}
\text{Det}\left( \underline{M}\right) =\prod\limits_{n=1}^{N}\left( \mu
_{n}\right) ~.~\square
\end{equation}

\textbf{Remark 2.1}. For fixed $q$ and $N,$ the $N$ eigenvalues $\mu _{n}$
of the $\left( N\times N\right) $-matrix $\underline{M}$ (see (\ref{MBis}))
depend only on the $r$ parameters $\alpha _{j}$ (see (\ref{mun})), while the
matrix $\underline{M}$ depends on the $s+r$ parameters $\beta _{k}$ and $%
\alpha _{j}$ via the dependence of the parameters $b_{k}$ respectively $%
a_{j} $ on $\beta _{k}$ respectively on $\alpha _{j}$ (see (\ref{bk})
respectively (\ref{aj})) and via the dependence of the $N$ zeros $\zeta _{n}$
on the parameters $\beta _{k}$ and $\alpha _{j}$. Hence the $\left( N\times
N\right) $-matrix $\underline{M}$ is \textit{isospectral} for variations of
the $s$ parameters $\beta _{k}$. And note moreover that the $N$ eigenvalues $%
\mu _{m}$ are \textit{rational} numbers if the $r$ parameters $\alpha _{j}$
are themselves \textit{rational} numbers, and $q$ is also a rational number:
a nontrivial \textit{Diophantine} property of the $\left( N\times N\right) $%
-matrix $\underline{M}$.~$\square $

\textbf{Remark 2.2}. All the above results are of course true as written
only provided the $N$ zeros $\zeta _{n}$ are all different among themselves;
but they remain valid by taking appropriate limits whenever this restriction
does not hold. $\square $

\textbf{Remark 2.3}. Immediate generalizations---whose explicit formulations
can be left to the interested reader--- of \textbf{Propositions 2.1} and 
\textbf{2.2}\textit{\ }obtain from these two propositions via the special
assignment $\alpha _{\hat{r}+p}=\beta _{\hat{s}+p}$ for $p=1,...,u$ with $u$
an arbitrary \textit{nonnegative integer} such that both $\hat{r}=r-u$ and $%
\hat{s}=s-u$ are \textit{positive integers}. These propositions refer then
to the $N$ zeros of the generalized basic hypergeometric polynomial $\tilde{P%
}_N\left( \alpha _{1},\ldots,\alpha _{\hat{r}};\beta _{1},\ldots,\beta _{%
\hat{s}};q;z\right)=P_{N}\left( \alpha _{1},\ldots,\alpha _{\hat{r}%
},\ldots,\alpha_r;\beta _{1},\ldots,\beta _{\hat{s}},\ldots,\beta_s;q;z%
\right) $ which depend---additionally to $N$ and $q$---only on the $\hat{r}+%
\hat{s}=r+s-2u$ parameters $\alpha _{j}$ with $j=1,...,\hat{r}=r-u$ and $%
\beta_{k}$ with $k=1,...,\hat{s}=s-u$, but feature quantities $b_{k}$ and $%
a_{j}$ (see (\ref{bk}) and (\ref{aj})) that depend on the $2+r+s$ parameters 
$N,$ $q,$ $\alpha _{j}$ with $j=1,...,r$ and $\beta_{k}$ with $k=1,...,s$. $%
\square $

Let us end this Section 2 by displaying explicitly the above results for
small values of the integers $r$, $s$ and $u$ (see \textbf{Remark 2.3}).

\subsection{The case $r=s=1,$ $u=0$}

If $r=s=1$ and $u=0$, the generalized basic hypergeometric polynomial~(\ref%
{PN}) is given by 
\end{subequations}
\begin{equation}
P_{N}(\alpha _{1};\beta _{1};q;z)=\sum_{m=0}^{N}\frac{(q^{-N};q)_{m}(\alpha
_{1};q)_{m}}{(q;q)_{m}(\beta _{1};q)_{m}}z^{m}.
\end{equation}%
Let $\zeta _{n},$ where $n=1,2,\ldots ,N$, be the zeros of $P_{N}(\alpha
_{1};\beta _{1};q;z)$. Then, by \textbf{Proposition 2.1}, 
\begin{eqnarray}
&&\Big( 1-\zeta_n q^{-N} +\frac{\beta_1}{q} -\alpha_1 \zeta_n \Big)%
\prod_{m=1}^{N}(\zeta _{n}q-\zeta _{m})  \notag \\
&&+\Big( -\frac{\beta_1}{q}+\alpha_1 \zeta_n q^{-N} \Big) %
\prod_{m=1}^{N}(\zeta _{n}q^2-\zeta _{m})=0 ~,~~n=1,2,\ldots ,N~.
\end{eqnarray}%
Define an $\left( N\times N\right) $-matrix $M$ componentwise as follows: 
\begin{subequations}
\label{Mter}
\begin{eqnarray}
&&M_{nn}=(q-1)^{2}g_{n}(1,\underline{\zeta }) \Big[ -1-\frac{\beta_1}{q}%
+\zeta_n (q^{-N} +\alpha_1) \Big] +(q^2-1)^{2}g_{n}(2,\underline{\zeta })%
\Big[\frac{\beta_1}{q}-\zeta_n \alpha_1 q^{-N} \Big]  \notag \\
&&+(q-1)f_{n}(1,\underline{\zeta })[-q^{-N}-\alpha_1]+(q^2-1)f_{n}(2,%
\underline{\zeta }) \alpha_1 q^{-N} ~, ~~n=1,2,...,N,  \label{MnnTer}
\end{eqnarray}%
\begin{eqnarray}
&&M_{nm}=\frac{\zeta _{n}}{(\zeta _{n}-\zeta _{m})^{2}}\Bigg\{ %
(q-1)^{2}f_{nm}(1,\underline{\zeta }) \Big[1+\frac{\beta_1}{q}%
-\zeta_n(q^{-N}+\alpha_1) \Big]  \notag \\
&&+(q^2-1)^2 f_{nm}(2,\underline{\zeta }) \Big[ -\frac{\beta_1}{q}+\zeta_n
\alpha_1 q^{-N} \Big] \Bigg\}~,~~ n,m=1,2,...,N,~~~n\neq m~,  \label{MnmTer}
\end{eqnarray}%
where the quantities $f_{nm}(p,\underline{\zeta })$ and $g_{n}(p,\underline{%
\zeta })$ are defined by~(\ref{fmngn}). By \textbf{Proposition 2.2}, the $N$
eigenvalues $\mu _{n}$ of this matrix are given by 
\end{subequations}
\begin{equation}
\mu _{n}=-(q^{-n}-1)(\alpha _{1}q^{N-n}-1),\;\;\;n=1,2,\ldots ,N.
\end{equation}%
By \textbf{Corollary 2.1}, the \textit{trace} of the matrix $\underline{M}$
is given by 
\begin{subequations}
\begin{equation}
\func{Tr}(\underline{M})=-\alpha _{1}\frac{q^{N+2}}{q^{2}-1}(1-q^{-2N-2})+%
\frac{q+\alpha _{1}q^{N+1}}{q-1}(1-q^{-N-1})-N-1
\end{equation}%
and the \textit{determinant} of $\underline{M}$ 
\begin{equation}
\det (\underline{M})=(-1)^{N}\prod_{n=1}^{N}\Big[(q^{-n}-1)(\alpha
_{1}q^{N-n}-1)\Big].
\end{equation}

\subsection{The case $r=2$, $s=1,$\ $u=0$}

If $r=2,s=1$ and $u=0$, the generalized basic hypergeometric polynomial~(\ref%
{PN}) is given by 
\end{subequations}
\begin{equation}
P_{N}(\alpha _{1},\alpha _{2};\beta _{1};q;z)=\sum_{m=0}^{N}\frac{%
(q^{-N};q)_{m}(\alpha _{1};q)_{m}(\alpha _{2};q)_{m}}{(q;q)_{m}(\beta
_{1};q)_{m}}(-1)^{m}q^{m(1-m)/2}z^{m}.
\end{equation}%
Let $\zeta _{n},$ where $n=1,2,\ldots ,N$, be the zeros of $P_{N}(\alpha
_{1},\alpha _{2};\beta _{1};q;z)$ and let $a_1=a_{1}\left( \underline{\alpha 
}\right) =\alpha _{1}+\alpha _{2}$, $a_2=a_{2}\left( \underline{\alpha }%
\right) =\alpha _{1}\alpha _{2}$ (see~(\ref{aj})). Then, by \textbf{%
Proposition 2.1}, 
\begin{eqnarray}
&&\left[ -1-\frac{\beta _{1}}{q}+\zeta _{n}(a_{2}+q^{-N}a_{1})\right]
\prod\limits_{m=1}^{N}\left( \zeta _{n}~q-\zeta _{m}\right) +\left[ \frac{%
\beta _{1}}{q}-\zeta _{n}q^{-N}a_{2}\right] \prod\limits_{m=1}^{N}\left(
\zeta _{n}~q^{2}-\zeta _{m}\right)  \notag \\
&&+\zeta _{n}\prod\limits_{m=1}^{N}\left( \zeta _{n}~q^{-1}-\zeta
_{m}\right) =0~,n=1,2,..,N~.
\end{eqnarray}

Moreover, by \textbf{Proposition 2.2}, the $N\times N$ matrix $\underline{M}$
defined componentwise by 
\begin{subequations}
\begin{eqnarray}
&&M_{nn}=(q-1)^{2}g_{n}(1,\underline{\zeta })\Big[ -1-\frac{\beta_1}{q}%
+\zeta_n(a_1 q^{-N}+a_2) \Big] +(q^2-1)^{2}g_{n}(2,\underline{\zeta })\Big[ 
\frac{\beta_1}{q}-\zeta_n a_2 q^{-N}\Big]  \notag \\
&&+(q^{-1}-1)^{2}g_{n}(-1,\underline{\zeta })\zeta_n-(q^{-1}-1)f_{n}(-1,%
\underline{\zeta })  \notag \\
&&+(q-1)f_{n}(1,\underline{\zeta })(-a_1 q^{-N}-a_2)+(q^{2}-1)f_{n}(2,%
\underline{\zeta }) a_2 q^{-N}~, ~~n=1,2,\ldots ,N,  \label{Mnnp2s1}
\end{eqnarray}%
\begin{eqnarray}
&&M_{nm}=\frac{\zeta _{n}}{(\zeta _{n}-\zeta _{m})^{2}}\Bigg\{ %
(q-1)^{2}f_{nm}(1,\underline{\zeta })\Big[1+\frac{\beta_1}{q}-\zeta_n(a_1
q^{-N}+a_2) \Big]  \notag \\
&&+(q^2-1)^{2}f_{nm}(2,\underline{\zeta })\Big[ -\frac{\beta_1}{q}+\zeta_n
a_2 q^{-N} \Big]  \notag \\
&&-(q^{-1}-1)^{2}f_{nm}(-1,\underline{\zeta })\zeta_n \Bigg\}%
~,~~n,m=1,2,\ldots ,N,n\neq m,  \label{Mnmp2s1}
\end{eqnarray}%
has the eigenvalues 
\end{subequations}
\begin{equation}
\mu _{n}=-q^{-N+n}(q^{-n}-1)(\alpha _{1}q^{N-n}-1)(\alpha
_{2}q^{N-n}-1),\;\;n=1,2,\ldots ,N.
\end{equation}%
By \textbf{Corollary 2.1}, the \textit{trace} of the matrix $\underline{M}$
is given by 
\begin{subequations}
\begin{equation}
\func{Tr}(\underline{M})=\frac{q^{-N}}{q^{2}-1}\Big\{-N(q^{2}-1)\Big[%
1+q^{N}(\alpha _{1}+\alpha _{2})\Big]+(q^{N}-1)\Big[q^{2}+\alpha _{1}+\alpha
_{2}-\alpha _{1}\alpha _{2}+q^{1+N}\alpha _{1}\alpha _{2}+q(1+\alpha
_{1}+\alpha _{2})\Big]\Big\}
\end{equation}%
and the \textit{determinant} of $\underline{M}$ 
\begin{equation}
\det (\underline{M})=(-1)^{N}\prod_{n=1}^{N}\Big[q^{-N+n}(q^{-n}-1)(\alpha
_{1}q^{N-n}-1)(\alpha _{2}q^{N-n}-1)\Big].
\end{equation}

\subsection{The case $r=s=2,$ $u=0$}

If $r=s=2$ and $u=0$, the generalized basic hypergeometric polynomial~(\ref%
{PN}) is given by 
\end{subequations}
\begin{equation}
P_{N}(\alpha _{1},\alpha _{2};\beta _{1}, \beta_2;q;z)=\sum_{m=0}^{N}\frac{%
(q^{-N};q)_{m}(\alpha _{1};q)_{m}(\alpha _{2};q)_{m}}{(q;q)_{m}(\beta
_{1};q)_{m}(\beta_{2};q)_{m}} z^{m}.
\end{equation}%
Let $\zeta _{n},$ where $n=1,2,\ldots ,N$, be the zeros of $P_{N}(\alpha
_{1},\alpha _{2};\beta _{1}, \beta_2;q;z)$ and let $a_1=a_{1}\left( 
\underline{\alpha }\right) =\alpha _{1}+\alpha _{2}$, $a_2=a_{2}\left( 
\underline{\alpha }\right) =\alpha _{1}\alpha _{2}$, (see~(\ref{aj})), $%
b_1=b_{1}\left( \underline{\beta }\right) =\beta_{1}+\beta _{2}$, $%
b_2=b_{2}\left( \underline{\beta }\right) =\beta _{1}\beta _{2}$ (see~(\ref%
{bk})). Then, by \textbf{Proposition 2.1}, 
\begin{eqnarray}
&&\left[ -1-\frac{b_1 }{q}+\zeta_n (q^{-N}+a_1)\right] \prod%
\limits_{m=1}^{N}\left( \zeta _{n}~q-\zeta _{m}\right)  \notag \\
&& +\left[ \frac{b _{1} }{q}+\frac{b_2 }{q^2}-\zeta _{n}(q^{-N} a_1+a_2 )%
\right] \prod\limits_{m=1}^{N}\left( \zeta _{n}~q^{2}-\zeta _{m}\right) 
\notag \\
&&+\left[ -\frac{b_2 }{q^2}+\zeta_n q^{-N} a_2 \right] \prod%
\limits_{m=1}^{N}\left( \zeta _{n}~q^{3}-\zeta _{m}\right) =0~,n=1,2,..,N~.
\label{prop21p2s2}
\end{eqnarray}

Moreover, by \textbf{Proposition 2.2}, the $N \times N$ matrix $\underline{M}
$ defined component-wise by 
\begin{subequations}
\begin{eqnarray}
&&M_{nn}=(q-1)^2 g_n(1,\underline{\zeta})\left[ 1+\frac{b_1}{q} -\zeta_n
(q^{-N}+ a_1) \right]  \notag \\
&& +(q^2-1)^2 g_n(2,\underline{\zeta})\left[ -\frac{b_1 }{q} -\frac{b_2}{q^2}%
+\zeta_n (q^{-N} a_1+a_2) \right]  \notag \\
&&+(q^3-1)^2 g_n(3,\underline{\zeta})\left[ \frac{b_2 }{q^2}-\zeta_n a_2
q^{-N} \right] +(q-1)f_n(1,\underline{\zeta})\left[ q^{-N}+a_1 \right] 
\notag \\
&& + (q^2-1)f_n(2,\underline{\zeta})\left[ -a_1 q^{-N}-a_2\right]
+(q^3-1)f_n(3,\underline{\zeta})a_2 q^{-N} , \; n=1,2,\ldots,N,
\label{Mnnp2s2}
\end{eqnarray}
\begin{eqnarray}
&&M_{nm}=\frac{\zeta_n}{(\zeta_n-\zeta_m)^2}\Bigg\{ (q-1)^2 f_{nm}(1,%
\underline{\zeta})\Big[ -1-\frac{b_1}{q}+\zeta_n (q^{-N}+a_1) \Big]  \notag
\\
&&+ (q^2-1)^2 f_{nm}(2,\underline{\zeta})\Big[ \frac{b_1}{q}+\frac{b_2}{q^2}%
-\zeta_n(a_1 q^{-N}+a_2) \Big]  \notag \\
&&+(q^3-1)^2 f_{nm}(3,\underline{\zeta})\Big[ -\frac{b_2}{q^2}+\zeta_n a_2
q^{-N} \Big] \Bigg\}, \;\; n,m=1,2,\ldots, N, n\neq m,  \label{Mnmp2s2}
\end{eqnarray}
has the eigenvalues 
\end{subequations}
\begin{equation}
\mu_n=-(q^{-n}-1)(\alpha_1 q^{N-n}-1)(\alpha_2 q^{N-n}-1), \;\;
n=1,2,\ldots,N.
\end{equation}
By \textbf{Corollary 2.1}, the \textit{trace} of the matrix $M$ is given by 
\begin{subequations}
\begin{equation}
\func{Tr}(M)=-\sum_{n=1}^N\Big[ (q^{-n}-1)(\alpha_1 q^{N-n}-1)(\alpha_2
q^{N-n}-1) \Big]
\end{equation}%
and the \textit{determinant} of $M$ 
\begin{equation}
\det (M)=(-1)^{N}\prod_{n=1}^{N}\Big[(q^{-n}-1)(\alpha_1 q^{N-n}-1)(\alpha_2
q^{N-n}-1)\Big].
\end{equation}

\subsection{The case $r=s=2$, $u=1$}

If $r=s=2$ and $u=1$, then $\beta _{2}=\alpha _{2}$ and the generalized
basic hypergeometric polynomial~(\ref{PN}) is given by 
\end{subequations}
\begin{equation}
\tilde{P}_N(\alpha _{1},\alpha _{2};\beta _{1};q;z)=P_{N}(\alpha _{1},\alpha
_{2};\beta _{1},\alpha _{2};q;z)=\sum_{m=0}^{N}\frac{(q^{-N};q)_{m}(\alpha
_{1};q)_{m}}{(q;q)_{m}(\beta _{1};q)_{m}}z^{m}.
\end{equation}%
Let $\zeta _{n},$ where $n=1,2,\ldots ,N$, be the zeros of $\tilde{P}%
_N(\alpha _{1},\alpha _{2};\beta _{1};q;z)$ and let $a_1=\alpha _{1}+\alpha
_{2}$, $a_2=\alpha _{1}\alpha _{2}$, (see~(\ref{aj})), $b_1=\beta
_{1}+\alpha _{2}$, $b_2=\beta _{1}\alpha _{2}$ (see~(\ref{bk})). Then, by 
\textbf{Proposition 2.1} and \textbf{Corollary 2.1}, the zeros $\underline{%
\zeta }$ satisfy algebraic relations~(\ref{prop21p2s2}) and the matrix $%
\underline{M}$ defined in terms of the zeros $\underline{\zeta }$ by
formulas~(\ref{Mp2s2}) has the eigenvalues 
\begin{equation}
\mu _{n}=-(q^{-n}-1)(\alpha _{1}q^{N-n}-1)(\alpha
_{2}q^{N-n}-1),\;\;n=1,2,\ldots ,N,
\end{equation}%
the \textit{trace} 
\begin{subequations}
\begin{equation}
\func{Tr}(\underline{M})=-\sum_{n=1}^{N}\Big[(q^{-n}-1)(\alpha
_{1}q^{N-n}-1)(\alpha _{2}q^{N-n}-1)\Big]
\end{equation}%
and the \textit{determinant} 
\begin{equation}
\det (\underline{M})=(-1)^{N}\prod_{n=1}^{N}\Big[(q^{-n}-1)(\alpha
_{1}q^{N-n}-1)(\alpha _{2}q^{N-n}-1)\Big].
\end{equation}

\bigskip

\section{Proof of \textbf{Proposition 2.2}}

Let the $t$-dependent \textit{monic} polynomial, of degree $N$ in $z$, be
characterized by its $N$ zeros $z_{n}\left( t\right) $, 
\end{subequations}
\begin{subequations}
\label{psixt}
\begin{equation}
\psi _{N}\left( z,t\right) =\tprod\limits_{n=1}^{N}\left[ z-z_{n}\left(
t\right) \right] ~,  \label{psixt1}
\end{equation}%
and by the $N$ coefficients $c_{m}\left( t\right) $ of its expansion in
powers of $z,$%
\begin{equation}
\psi _{N}\left( z,t\right) =z^{N}+\sum_{m=1}^{N}\left[ c_{m}\left( t\right)
~z^{N-m}\right] ~.  \label{psiNc}
\end{equation}%
It is plain that these two representations, (\ref{psixt1}) and (\ref{psiNc}%
), are consistent, implying a uniquely defined expression of the $N$
coefficients $c_{m}\left( t\right) $ in terms of the $N$ zeros $z_{n}\left(
t\right) $, and an expression---unique up to permutations and of course
explicitly known only for $N\leq 4$---of the $N$ zeros $z_{n}\left( t\right) 
$ in terms of the $N$ coefficients $c_{m}\left( t\right) $.

The next step is to assume that this $t$-dependent polynomial $\psi
_{N}\left( z,t\right) $ satisfy the following \textit{linear} Differential 
\textit{q-}Difference Equation (D\textit{q}DE): 
\end{subequations}
\begin{eqnarray}
&&\frac{\partial ~\psi _{N}\left( z,t\right) }{\partial ~t}=z^{-1}~\left[
\Delta _{1}~\dprod\limits_{k=1}^{s}\left( \Delta _{\beta _{k}/q}\right) %
\right] ~\psi _{N}\left( z,t\right)  \notag  \label{DqDE} \\
&&-\left[ \Delta _{q^{-N}}\dprod\limits_{j=1}^{r}\left( \Delta _{\alpha
_{j}}\right) \right] ~\psi _{N}\left( q^{s-r}~z,t\right) ~,
\end{eqnarray}%
where the operator $\Delta _{\gamma }$, acting on functions of the variable $%
z$, is defined as follows: 
\begin{subequations}
\begin{equation}
\Delta _{\gamma }=\left( \gamma ~\delta -1\right) ~~~\text{with~~~}\delta
~f\left( z\right) =f\left( q~z\right)  \label{Deltaa}
\end{equation}%
(see (\ref{deltas})). Note that this implies%
\begin{equation}
\Delta _{\gamma }~z^{p}=\left( \gamma q^{p}-1\right) ~z^{p}~.  \label{Deltab}
\end{equation}

Let us first of all check whether this is consistent with the fact that $%
\psi _{N}\left( z,t\right) $ is a polynomial of degree $N$ in $z$. Since the
operator $\Delta _{\gamma },$ when applied to a power of $z,$ does not
change that power but only multiplies it by a number, see~(\ref{Deltab}), it
is plain that this is guaranteed by the fact that $\Delta
_{q^{-N}}~z^{N}=\left( 1-1\right) ~z^{N}=0$ (see (\ref{Deltab})); while of
course clearly $\Delta _{1}c=0$ (where $c$ indicates any constant, i. e. a
quantity independent of the variable $z$).

Next, let us see what the D\textit{q}DE (\ref{DqDE}) implies for the
coefficients $c_{m}\left( t\right) ,$ see (\ref{psiNc}). Via (\ref{psiNc})
and (\ref{Deltab}) it clearly amounts to the following \textit{autonomous
linear} system of ODEs: 
\end{subequations}
\begin{subequations}
\label{cmt}
\begin{eqnarray}
&&\dot{c}_{m}\left( t\right) =\left[ \left( q^{N-m+1}-1\right)
~\dprod\limits_{k=1}^{s}\left( \beta _{k}~q^{N-m}-1\right) \right]
~c_{m-1}\left( t\right)  \notag \\
&&-\left[ q^{\left( s-r\right) \left( N-m\right) }\left( q^{-m}-1\right)
~\dprod\limits_{j=1}^{r}\left( \alpha _{j}~q^{N-m}-1\right) \right]
~c_{m}\left( t\right) ~,  \notag \\
&&m=1,2,...,N~,~~~\text{with}~~~c_{0}=1~.  \label{cmdot}
\end{eqnarray}%
Of course here and below a superimposed dot denotes a $t$-differentiation.

Note the formal consistency of this system of evolution equations with the
assignment $c_{0}=1,$ hence with the fact that $\psi _{N}\left( z,t\right) $
is a \textit{monic} polynomial of degree $N$ in $z$, see (\ref{psiNc}).

It is plain that the solution of this system reads%
\begin{equation}
c_{m}\left( t\right) =\sum_{n=0}^{N}\left[ \eta _{n}~u_{n}^{\left( m\right)
}\exp \left( \mu _{m}~t~\right) \right] ~,
\end{equation}%
where the $N$-vectors $\underline{u}^{\left( m\right) }$, respectively the $%
N $ numbers $\mu _{n}$, are the $N$ eigenvectors, respectively the
corresponding $N$ eigenvalues, of the $\left( N\times N\right) $-matrix $%
\underline{C}$\ with elements%
\begin{equation}
C_{nn}=-q^{\left( s-r\right) \left( N-n\right) }\left( q^{-n}-1\right)
~\dprod\limits_{j=1}^{r}\left( \alpha _{j}~q^{N-n}-1\right)
\,,~~~n=1,2,...,N~,  \label{Cnn}
\end{equation}%
\begin{equation}
C_{n,n-1}=\left( q^{N-n+1}-1\right) ~\dprod\limits_{k=1}^{s}\left( \beta
_{k}~q^{N-n}-1\right) \,,~~~n=2,...,N~,  \label{Cnm}
\end{equation}%
and all other elements vanishing: 
\begin{equation}
\underline{C}~\underline{u}^{\left( m\right) }=\mu _{n}~\underline{u}%
^{\left( m\right) }~;
\end{equation}%
while the $N$ ($t$-independent) numbers $\eta _{n}$ can be arbitrarily
assigned---getting thereby the \textit{general solution} of system (\ref%
{cmdot})---or can be adjusted to fit the initial data $c_{m}\left( 0\right) $
so that%
\begin{equation}
c_{m}\left( 0\right) =\sum_{n=1}^{N}\left[ \eta _{n}~u_{n}^{\left( m\right) }%
\right] ~,~~~m=1,...,N
\end{equation}%
---getting thereby the solution of the \textit{initial-value problem} for
system (\ref{cmdot}). The triangular character of the $\left( N\times
N\right) $-matrix $\underline{C},$ see (\ref{Cnn}) and (\ref{Cnm}), implies
that its $N$ eigenvalues $\mu _{n}$ coincide with its diagonal elements:%
\begin{equation}
\mu _{n}=-q^{\left( s-r\right) \left( N-n\right) }\left( q^{-n}-1\right)
~\dprod\limits_{j=1}^{r}\left( \alpha _{j}~q^{N-n}-1\right)
\,,~~~n=1,2,...,N~.  \label{eigenvaluesmun}
\end{equation}

Our next task is to discuss the $t$-evolution of the $N$ zeros $z_{n}\left(
t\right) $ of $\psi \left( z,t\right) ,$ see (\ref{zerosofpsit}), implied by
the D\textit{q}DE (\ref{DqDE}).

The first observation is that the equilibrium---i. e., $t$%
-independent---solution $\bar{\psi}\left( z\right) $ of D\textit{q}DE (\ref%
{DqDE}) is the generalized basic hypergeometric polynomial $p_{N}\left(
z\right)=p_N(\underline{\alpha}; \underline{\beta};q;z) $, see (\ref{pN}), 
\end{subequations}
\begin{equation}
\bar{\psi}\left( z\right) =p_{N}\left( z\right) ~.
\end{equation}%
This is implied by the $q$-difference equation satisfied by $p_{N}\left(
z\right) $, see (\ref{qDE}), which clearly implies that the right-hand side
of the D\textit{q}DE (\ref{DqDE}) vanishes for $\psi \left( z,t\right) =\bar{%
\psi}\left( z\right) =p_{N}\left( z\right) $ (and note the consistency
implied by the fact that $\bar{\psi}\left( z\right) $ and $p_{N}\left(
z\right) $ are both \textit{monic} polynomials of degree $N$). Hence the
equilibrium---i. e., $t$-independent---configuration of the $N$ zeros $%
z_{n}\left( t\right) $ of $\psi \left( z,t\right) ,$ see (\ref{zerosofpsit}%
), is%
\begin{equation}
z_{n}\left( t\right) =z_{n}\left( 0\right) =\bar{z}_{n}=\zeta
_{n}~,~~~n=1,2,...,N~,  \label{Equizn}
\end{equation}%
where the $N$ numbers $\zeta _{n}$ are the $N$ zeros of the (monic)
generalized basic hypergeometric polynomial $p_{N}\left( z\right) $, see (%
\ref{Zeros}).

Next, let us reformulate D\textit{q}DE (\ref{DqDE}) as follows: 
\begin{subequations}
\label{Eqpsi}
\begin{equation}
\left( \frac{\partial }{\partial ~t}\right) ~\psi _{N}\left( z,t\right)
=RHS\left( z,t\right) ~,  \label{Eqpsi1}
\end{equation}%
\begin{eqnarray}
&&RHS\left( z,t\right) =(-1)^{s}z^{-1}\Bigg\{\psi _{N}(qz,t)-\psi
_{N}(z,t)+\sum_{k=1}^{s}b_{k}(\underline{\beta })\frac{(-1)^{k}}{q^{k}}\left[
\psi _{N}(q^{k+1}z,t)-\psi _{N}(q^{k}z,t)\right] \Bigg\}  \notag \\
&&-(-1)^{r}\Bigg\{q^{-N}\psi _{N}(q^{s-r+1}z,t)-\psi _{N}(q^{s-r}z,t)  \notag
\\
&&+\sum_{j=1}^{r}a_{j}(\underline{\alpha })(-1)^{j}\left[ q^{-N}\psi
_{N}(q^{j+s+1-r}z,t)-\psi _{N}(q^{j+s-r}z,t)\right] \Bigg\}~,  \label{RHS}
\end{eqnarray}%
by repeating, on the right-hand side of D\textit{q}DE (\ref{DqDE}), the same
development that led, in the Appendix, from (\ref{Ex131}) to (\ref{qdifferEq}%
); this of course implies that the quantities $b_{k}\left( \underline{\beta }%
\right) $ respectively $a_{j}\left( \underline{\alpha }\right) $ are, here
and hereafter, defined as in the Appendix in terms of the parameters $\beta
_{k}$ respectively $\alpha _{j}$, see (\ref{bk}) and (\ref{aj}).

Next, let us insert in this D\textit{q}DE, (\ref{Eqpsi}), the representation
(\ref{psixt1}) of the monic polynomial $\psi \left( z,t\right) $ via its
zeros $z_{n}\left( t\right) $. The left-hand side of this D\textit{q}DE then
reads (by logarithmic $t$-differentiation of (\ref{psixt1})) 
\end{subequations}
\begin{subequations}
\begin{equation}
\left( \frac{\partial }{\partial ~t}\right) ~\psi _{N}\left( z,t\right)
=-\psi _{N}\left( z,t\right) ~\sum_{m=1}^{N}\left[ \frac{\dot{z}_{m}\left(
t\right) }{z-z_{m}\left( t\right) }\right] =-\sum_{m=1}^{N}\left\{ \dot{z}%
_{m}\left( t\right) ~\prod\limits_{\ell =1,~\ell \neq m}^{N}\left[ z-z_{\ell
}\left( t\right) \right] \right\} ~,  \label{psiNdot}
\end{equation}%
implying, for $z=z_{n}\left( t\right) $,%
\begin{equation}
\left. \left( \frac{\partial }{\partial ~t}\right) ~\psi _{N}\left(
z,t\right) \right\vert _{z=z_{n}\left( t\right) }=-\dot{z}_{n}\left(
t\right) ~\prod\limits_{\ell =1,~\ell \neq n}^{N}\left[ z_{n}\left( t\right)
-z_{\ell }\left( t\right) \right] ~.
\end{equation}%
While the right-hand side (\ref{RHS}) of (\ref{Eqpsi1}) clearly reads 
\end{subequations}
\begin{subequations}
\begin{eqnarray}
&&RHS\left( z,t\right) =(-1)^{s}z^{-1}\Bigg\{\prod_{\ell =1}^{N}\left[
qz-z_{\ell }(t)\right] -\prod_{\ell =1}^{N}\left[ z-z_{\ell }(t)\right] 
\notag \\
&&+\sum_{k=1}^{s}b_{k}(\underline{\beta })\frac{(-1)^{k}}{q^{k}}\left[
\prod_{\ell =1}^{N}\left[ q^{k+1}z-z_{\ell }(t)\right] -\prod_{\ell =1}^{N}%
\left[ q^{k}z-z_{\ell }(t)\right] \right] \Bigg\}  \notag \\
&&-(-1)^{r}\Bigg\{q^{-N}\prod_{\ell =1}^{N}\left[ q^{s-r+1}z-z_{\ell }(t)%
\right] -\prod_{\ell =1}^{N}\left[ q^{s-r}z-z_{\ell }(t)\right]  \notag \\
&&+\sum_{j=1}^{r}a_{j}(\underline{\alpha })(-1)^{j}\left[ q^{-N}\prod_{\ell
=1}^{N}\left[ q^{j+s+1-r}z-z_{\ell }(t)\right] -\prod_{\ell
=1}^{N}[q^{j+s-r}z-z_{\ell }(t))]\right] \Bigg\}~,
\end{eqnarray}%
implying, for $z=z_{n}\left( t\right) $, 
\begin{eqnarray}
&&RHS\left( z_{n}\left( t\right) ,t\right) =(-1)^{s}z_{n}(t)^{-1}\Bigg\{%
\prod_{\ell =1}^{N}\left[ qz_{n}(t)-z_{\ell }(t)\right]  \notag \\
&&+\sum_{k=1}^{s}b_{k}(\underline{\beta })\frac{(-1)^{k}}{q^{k}}\left[
\prod_{\ell =1}^{N}\left[ q^{k+1}z_{n}(t)-z_{\ell }(t)\right] -\prod_{\ell
=1}^{N}\left[ q^{k}z_{n}(t)-z_{\ell }(t)\right] \right] \Bigg\}  \notag \\
&&-(-1)^{r}\Bigg\{q^{-N}\prod_{\ell =1}^{N}\left[ q^{s-r+1}z_{n}(t)-z_{\ell
}(t)\right] -\prod_{\ell =1}^{N}\left[ q^{s-r}z_{n}(t)-z_{\ell }(t)\right] 
\notag \\
&&+\sum_{j=1}^{r}a_{j}(\underline{\alpha })(-1)^{j}\left[ q^{-N}\prod_{\ell
=1}^{N}\left[ q^{j+s+1-r}z_{n}(t)-z_{\ell }(t)\right] -\prod_{\ell =1}^{N}%
\left[ q^{j+s-r}z_{n}(t)-z_{\ell }(t)\right] \right] \Bigg\}~.
\end{eqnarray}

It is thus seen that the equations of motion characterizing the $t$%
-evolution of the $N$ zeros $z_{n}\left( t\right) $ of $\psi \left(
z,t\right) $ read as follows (of course below a superimposed dot denotes a $%
t $-differentiation, and we omit for notational simplicity to display the $t$%
-dependence of the zeros): 
\end{subequations}
\begin{subequations}
\begin{eqnarray}
&&\dot{z}_{n}=(-1)^{s+1}\Bigg\{(q-1)f_{n}(1,\underline{z})+%
\sum_{k=1}^{s}b_{k}(\underline{\beta })\frac{(-1)^{k}}{q^{k}}\Big[%
(q^{k+1}-1)f_{n}(k+1,\underline{z})-(q^{k}-1)f_{n}(k,\underline{z})\Big]%
\Bigg\}  \notag \\
&&+(-1)^{r}z_{n}\Bigg\{q^{-N}(q^{s-r+1}-1)f_{n}(s-r+1,\underline{z}%
)-(q^{s-r}-1)f_{n}(s-r,\underline{z})  \notag \\
&&+\sum_{j=1}^{r}a_{j}(\underline{\alpha })(-1)^{j}\Big[%
q^{-N}(q^{j+s+1-r}-1)f_{n}(j+s+1-r,\underline{z})-(q^{j+s-r}-1)f_{n}(j+s-r,%
\underline{z})\Big]\Bigg\}~,  \label{zerosofpsit}
\end{eqnarray}%
where 
\begin{equation}
f_{n}(p,\underline{z})=f_{n}(p,z_{1},\ldots ,z_{N})=\prod_{\ell =1,\ell \neq
n}^{N}\left( \frac{q^{p}~z_{n}-z_{\ell }}{z_{n}-z_{\ell }}\right)
,\;n=1,2,...,N~.  \label{fnpz}
\end{equation}

This is an interesting dynamical system, a complete investigation of which
is beyond the scope of the present paper.\ But before proceeding with our
task, let us pause and recall that the first idea to relate the zeros of
polynomials to the equilibria of a dynamical system goes back to Stieltjes
and Sz\"{e}go \cite{S1885}, was resuscitated in \cite{C1978} to identify
\textquotedblleft solvable\textquotedblright\ many-body problems (see also
the extended treatment of this approach in \cite{C2001}), and then
extensively used to obtain results concerning the zeros of the classical
polynomials and of Bessel functions, see the paper \cite{ABCOP1979} where
several such findings are derived and reviewed. For more recent developments
along somewhat analogous lines see, for instance, \cite{OS2004}, \cite%
{vD2005}, \cite{GVZ2014}, \cite{C2014}, \cite{S2014}.

Here we need to focus only on the behavior of this dynamical system, (\ref%
{zerosofpsit}), in the immediate neighborhood of its equilibrium
configuration $z_{n}\left( t\right) =\bar{z}_{n}=\zeta _{n}$, see (\ref%
{Equizn}). To this end we set 
\end{subequations}
\begin{subequations}
\label{Ansatz}
\begin{equation}
z_{n}\left( t\right) =\zeta _{n}+\varepsilon ~\xi _{n}\left( t\right)
~,~~~n=1,..,N~,
\end{equation}%
implying 
\begin{equation}
\dot{z}_{n}\left( t\right) =\varepsilon ~\dot{\xi}_{n}\left( t\right)
~,~~~n=1,..,N~,
\end{equation}%
with $\varepsilon $ infinitesimal. To aid the task of linearizing system~(%
\ref{zerosofpsit}), we introduce the quantities 
\end{subequations}
\begin{subequations}
\label{fmngn}
\begin{equation}
f_{nm}(p,\underline{z})=f_{nm}(p,z_{1},\ldots ,z_{N})=\prod_{\ell =1,\ell
\neq n,m}^{N}\left( \frac{q^{p}~z_{n}-z_{\ell }}{z_{n}-z_{\ell }}\right)
\label{fnmpz}
\end{equation}%
(implying $f_{nm}\left( p,\underline{z}\right) =f_{n}(p,\underline{z})\left[
\left( z_{n}-z_{m}\right) /\left( q^{p}~z_{n}-z_{m}\right) \right] $; see (%
\ref{fnpz})), and 
\begin{equation}
g_{n}(p,\underline{z})=g_{n}(p,z_{1},\ldots ,z_{N})=\sum_{k=1,k\neq n}^{N}%
\left[ f_{nk}(p,\underline{z})~\frac{z_{k}}{(z_{n}-z_{k})^{2}}\right] ~;
\label{gnpz}
\end{equation}%
and we note that 
\begin{equation}
\frac{\partial f_{n}}{\partial z_{n}}(p,\underline{z})=(1-q^{p})~g_{n}(p,%
\underline{z})  \label{dfndzn}
\end{equation}%
and 
\begin{equation}
\frac{\partial f_{n}}{\partial z_{m}}(p,\underline{z})=(q^{p}-1)~f_{nm}(p,%
\underline{z})~\frac{z_{n}}{(z_{n}-z_{m})^{2}}~,  \label{dfndzm}
\end{equation}%
where $n,m=1,2,\ldots ,N$ and $n\neq m$.

The insertion of \textit{ansatz}~(\ref{Ansatz}) in the equations of motion (%
\ref{zerosofpsit}) is, to order $\varepsilon ^{0}=1$, clearly consistent via 
\textbf{Proposition 2.1}. To order $\varepsilon $, it yields the \textit{%
linearized} system of ODEs 
\end{subequations}
\begin{equation}
\underline{\dot{\xi}}=\underline{M}~\underline{\xi }~;~~~\dot{\xi}%
_{n}=\sum_{m=1}^{N}\left( M_{nm}~\xi _{m}\right) ~,~~~n=1,...,N~,
\label{Eqxn}
\end{equation}%
with the $\left( N\times N\right) $-matrix $\underline{M}$ defined,
componentwise, by (\ref{MBis}) (the diligent reader might wish to check,
using the notation and the formulas in~(\ref{fnpz}) and~(\ref{fmngn}), the
relevant computation, which is standard but somewhat cumbersome---although
certainly doable without the help of a computer). Here and hereafter the
notation $~\underline{\xi }\equiv ~\underline{\xi }\left( t\right) $ denotes
the $N$-vector of components $\xi _{n}\equiv \xi _{n}\left( t\right) $. As
for the implications to orders $\varepsilon ^{p}$ with $p=2,3,...$ of the
insertion of the \textit{ansatz} (\ref{Ansatz}) in the equations of motion (%
\ref{zerosofpsit}), they consist of additional systems of algebraic
equations satisfied by the zeros $\zeta _{n}$ of the generalized basic
hypergeometric polynomial $P_{N}\left( z\right) $ or $p_{N}\left( z\right) $%
, see (\ref{PN}), (\ref{pN}) and (\ref{Zeros}), the explicit display of
which is left to the interested reader (to get them it might be expedient to
take advantage of appropriate computer packages such as Mathematica or
Maple).

The \textit{general} solution of the system of \textit{linear} ODEs (\ref%
{Eqxn}) reads of course as follows: 
\begin{equation}
\underline{\xi }\left( t\right) =\sum_{m=1}^{N}\left[ \eta _{m}~\exp \left( 
\tilde{\mu}_{m}~t\right) ~\underline{v}^{\left( m\right) }\right] ~,
\end{equation}%
where the $N$ ($t$-independent) parameters $\eta _{m}$ can be arbitrarily
assigned (or adjusted to satisfy the $N$ \textit{initial} conditions $\xi
_{n}\left( 0\right) $), while the numbers $\tilde{\mu}_{m}$ respectively the 
$N$-vectors $\underline{v}^{\left( m\right) }$ are clearly the $N$
eigenvalues respectively the $N$ eigenvectors of the matrix $\underline{M},$%
\begin{equation}
\underline{M}~\underline{v}^{\left( m\right) }=\tilde{\mu}_{m}~\underline{v}%
^{\left( m\right) }~,~\ \ m=1,...,N~.  \label{EigenM}
\end{equation}%
But the behavior of the dynamical system (\ref{Eqxn}) in the \textit{%
immediate vicinity} of its \textit{equilibria} cannot differ from its 
\textit{general} behavior, which is characterized by the $N$ exponentials $%
\exp \left( \mu _{m}~t\right) $, as implied by the relation between the $N$
zeros $z_{n}\left( t\right) $ and the coefficients $c_{m}\left( t\right) $
of the monic polynomial (of degree $N$ in $z$) $\psi _{N}\left( z,t\right) $%
, see (\ref{psixt}), and by the explicit formula, see (\ref{cmt}), detailing
the $t$-evolution of the $N$ coefficients $c_{m}\left( t\right) .$ Hence the
(set of) eigenvalues $\tilde{\mu}_{m}$ of the matrix $\underline{M}$, see (%
\ref{MBis}), must coincide with the (set of) eigenvalues $\mu _{m}$, see (%
\ref{eigenvaluesmun}), of the matrix $\underline{C}$, see (\ref{cmt}). 
\textbf{Proposition 2.2} is thereby proven.

\bigskip

\section{Outlook}

A follow-up to the present paper might explore the dynamical system (\ref%
{zerosofpsit}), as well as other dynamical systems obtained by analogous
methods, also connected with hypergeometric or basic hypergeometric
polynomials, but featuring more interesting equations of motion, for
instance of Newtonian type---allowing their interpretation as ``many-body
models''.

Another possible direction of further investigation might try and extend the
approach and findings, reported in this paper for the $N$\ zeros of basic
hypergeometric polynomials of order $N$, to the, generally infinite, set of
zeros of (nonpolynomial) generalized basic hypergeometric functions.

\bigskip

\section{Acknowledgements}

One of us (OB) would like to acknowledge with thanks the hospitality of the
Physics Department of the University of Rome \textquotedblleft La Sapienza''
on the occasion of three two-week visits there in June 2012, May 2013 and
June-July 2014; this paper was initiated during the last of these visits.
The other one (FC) would like to acknowledge with thanks the hospitality of
Concordia College for a one-week visit there in November 2013.

\bigskip

\section{Appendix A: Properties of the generalized basic hypergeometric
polynomials}

In this appendix we report some properties of the generalized basic
hypergeometric polynomial $p_{N}\left( z\right) \equiv p_{N}\left( 
\underline{\alpha };\underline{\beta };q;z\right) $ (see (\ref{PN}) and (\ref%
{pN}) and \textbf{Notation 1.1}); and we prove \textbf{Proposition 2.1}. It
is understood that, throughout this Appendix, the $r+s+2$ parameters $\alpha
_{j}$, $\beta _{k}$, $q\neq 1$ and $N$ have fixed values; the indication of
the dependence upon them is omitted whenever this simplifies the notation at
no cost in terms of clarity.

Our starting point is the following $q$-difference equation satisfied by the
generalized basic hypergeometric function $_{r+1}\phi _{s}\left( \alpha
_{0},\alpha _{1,}...,\alpha _{r};\beta _{1},...,\beta _{s};z\right) $ (see
Exercise 1.31 on page 27 of \cite{GR1990}, and note the replacement of $r$
with $r+1$ and of $\alpha _{0}$ with $q^{-N}$): 
\begin{subequations}
\label{DDE}
\begin{eqnarray}
&\Delta _{1}~\left[ \dprod\limits_{k=1}^{s}\left( \Delta _{\beta
_{k}/q}\right) \right] ~_{r+1}\phi _{s}\left( q^{-N},\alpha _{1,}...,\alpha
_{r};\beta _{1},...,\beta _{s};q;z\right) &  \notag \\
&-z~\Delta _{q^{-N}}~\left[ \dprod\limits_{j=1}^{r}\left( \Delta _{\alpha
_{j}}\right) \right] ~_{r+1}\phi _{s}\left( q^{-N},\alpha _{1,}...,\alpha
_{r};\beta _{1},...,\beta _{s};q;zq^{s-r}\right) =&0~,  \label{Ex131}
\end{eqnarray}%
where the difference operators $\Delta _{\gamma }$ and $\delta $ operate as
follows on functions $f\left( z\right) $ of the variable $z$: 
\begin{equation}
\Delta _{\gamma }~f\left( z\right) =\gamma f\left( zq\right) -f\left(
z\right) ~,~~~\Delta _{\gamma }=\gamma \delta -1~,~~~\delta ~f\left(
z\right) =f\left( zq\right) ~.  \label{deltas}
\end{equation}%
Note that the operators $\Delta _{\gamma _{1}}$ and $\Delta _{\gamma _{2}}$
commute,%
\begin{equation}
\Delta _{\gamma _{1}}\Delta _{\gamma _{2}}~f\left( z\right) =\gamma
_{1}\gamma _{2}~f\left( q^{2}z\right) -\left( \gamma _{1}+\gamma _{2}\right)
~f\left( qz\right) +f\left( z\right) =\Delta _{\gamma _{2}}\Delta _{\gamma
_{1}}~f\left( z\right) .
\end{equation}

Via (\ref{PN}) and (\ref{pN}) this implies that the generalized basic
hypergeometric polynomial of order $N$ satisfies the following $q$%
-difference equation: 
\end{subequations}
\begin{equation}
\Delta _{1}~\left[ \dprod\limits_{k=1}^{s}\left( \Delta _{\beta
_{k}/q}\right) \right] ~p_{N}\left( \underline{\alpha };\underline{\beta }%
;q;z\right) -z~\Delta _{q^{-N}}~\left[ \dprod\limits_{j=1}^{r}\left( \Delta
_{\alpha _{j}}\right) \right] ~p_{N}\left( \underline{\alpha };\underline{%
\beta };q;zq^{s-r}\right) =0~.  \label{qDE}
\end{equation}

It is now convenient to introduce the identities 
\begin{eqnarray}
&&\dprod\limits_{k=1}^{s}\left( \Delta _{\beta _{k}/q}\right)
=\dprod\limits_{k=1}^{s}\left( \beta _{k}\frac{\delta }{q}-1\right) =\left(
-1\right) ^{s}~\dprod\limits_{k=1}^{s}\left( 1-\beta _{k}\frac{\delta }{q}%
\right)  \notag \\
&=&\left( -1\right) ^{s}~\left\{ 1+\sum_{k=1}^{s}\left[ b_{k}\left( 
\underline{\beta }\right) ~\left( -\frac{\delta }{q}\right) ^{k}\right]
\right\} ~,
\end{eqnarray}%
where 
\begin{subequations}
\label{bk}
\begin{equation}
\dprod\limits_{k=1}^{s}\left( 1+\beta _{k}~x\right) \equiv 1+\sum_{k=1}^{s} 
\left[ b_{k}\left( \underline{\beta }\right) ~x^{k}\right] ~,
\end{equation}%
so that the quantities $b_{k}\left( \underline{\beta }\right) $ are defined
as follows,%
\begin{equation}
b_{1}\left( \underline{\beta }\right) =\sum_{k=1}^{s}\left( \beta
_{k}\right) ~,  \label{b1}
\end{equation}%
\begin{equation}
b_{2}\left( \underline{\beta }\right) =\sum_{k_{1},k_{2}=1,~k_{1}\neq
k_{2}}^{s}\left( \beta _{k_{1}}~\beta _{k_{2}}\right) ~,
\end{equation}%
\begin{equation}
b_{3}\left( \underline{\beta }\right) =\sum_{k_{1},k_{2},k_{3}=1,~k_{1}\neq
k_{2},~k_{2}\neq k_{3},~k_{3}\neq k_{1},}^{s}\left( \beta _{k_{1}}~\beta
_{k_{2}}~\beta _{k_{3}}\right) ~,
\end{equation}%
and so on, up to%
\begin{equation}
b_{s}\left( \underline{\beta }\right) =\dprod\limits_{k=1}^{s}\left( \beta
_{k}\right) ~;
\end{equation}%
as well as the analogous identities 
\end{subequations}
\begin{eqnarray}
&&\dprod\limits_{j=1}^{r}\left( \Delta _{\alpha _{j}}\right)
=\dprod\limits_{k=1}^{r}\left( \alpha _{j}\delta -1\right) =\left( -1\right)
^{r}~\dprod\limits_{j=1}^{r}\left( 1-\alpha _{j}\delta \right)  \notag \\
&=&\left( -1\right) ^{r}~\left\{ 1+\sum_{j=1}^{r}\left[ a_{j}\left( 
\underline{\alpha }\right) ~\left( -\delta \right) ^{j}\right] \right\} ~,
\end{eqnarray}%
where 
\begin{subequations}
\label{aj}
\begin{equation}
\dprod\limits_{j=1}^{r}\left( 1+\alpha _{j}~x\right) \equiv 1+\sum_{j=1}^{r} 
\left[ a_{j}\left( \underline{\alpha }\right) ~x^{j}\right] ~,
\end{equation}%
so that the quantities $a_{j}\left( \underline{\alpha }\right) $ are defined
as follows,%
\begin{equation}
a_{1}\left( \underline{\alpha }\right) =\sum_{k=1}^{r}\left( \alpha
_{j}\right) ~,  \label{a1}
\end{equation}%
\begin{equation}
a_{2}\left( \underline{\alpha }\right) =\sum_{j_{1},j_{2}=1,~j_{1}\neq
j_{2}}^{r}\left( \alpha _{j_{1}}~\alpha _{j_{2}}\right) ~,
\end{equation}%
\begin{equation}
a_{3}\left( \underline{\alpha }\right) =\sum_{j_{1},j_{2},j_{3}=1,~j_{1}\neq
j_{2},~j_{2}\neq j_{3},~j_{3}\neq j_{1},}^{r}\left( \alpha _{j_{1}}~\alpha
_{j_{2}}~\alpha _{j_{3}}\right) ~,
\end{equation}%
and so on, up to%
\begin{equation}
a_{r}\left( \underline{\beta }\right) =\dprod\limits_{j=1}^{r}\left( \alpha
_{j}\right) ~.
\end{equation}

Hence the $q$-difference equation (\ref{qDE}) satisfied by the \textit{%
generalized basic hypergeometric polynomial} $p_{N}\left( z\right) $ (see (%
\ref{pN}) and (\ref{PN})) reads as follows (see (\ref{deltas})): 
\end{subequations}
\begin{subequations}
\begin{eqnarray}
&\left( 1-\delta \right) ~\left\{ 1+\sum\limits_{k=1}^{s}\left[ b_{k}\left( 
\underline{\beta }\right) ~\left( -\frac{\delta }{q}\right) ^{k}\right]
\right\} ~p_{N}\left( z\right) &  \notag \\
&-z~\left( 1-q^{-N}\delta \right) ~\left( -1\right) ^{r-s}~\left\{
1+\sum\limits_{j=1}^{r}\left[ a_{j}\left( \underline{\alpha }\right) ~\left(
-\delta \right) ^{j}\right] \right\} ~p_{N}\left( z~q^{s-r}\right) =&0~,
\end{eqnarray}%
hence%
\begin{eqnarray}
&\left\{ 1-\delta +\sum\limits_{k=1}^{s}\left[ \left( -q\right)
^{-k}b_{k}\left( \underline{\beta }\right) ~\left( \delta ^{k}-\delta
^{k+1}\right) \right] \right\} ~p_{N}\left( z\right) &  \notag \\
&-\left( -1\right) ^{r-s}~z~\left\{ 1-q^{-N}\delta +\sum\limits_{j=1}^{r} 
\left[ \left( -1\right) ^{j}a_{j}\left( \underline{\alpha }\right) ~\left(
\delta ^{j}-q^{-N}\delta ^{j+1}\right) \right] \right\} ~p_{N}\left(
z~q^{s-r}\right) =&0~,
\end{eqnarray}%
hence 
\begin{eqnarray}
p_{N}\left( z\right) -p_{N}\left( z~q\right) +\sum_{k=1}^{s}\left( -q\right)
^{-k}b_{k}\left( \underline{\beta }\right) ~\left[ p_{N}\left(
z~q^{k}\right) -p_{N}\left( z~q^{k+1}\right) \right] &&  \notag \\
-\left( -1\right) ^{r-s}~z~\left[ p_{N}\left( z~q^{s-r}\right)
-q^{-N}p_{N}\left( z~q^{s-r+1}\right) \right] &&  \notag \\
-\left( -1\right) ^{r-s}~z~\left\{ \sum_{j=1}^{r}\left[ \left( -1\right)
^{j}a_{j}\left( \underline{\alpha }\right) ~p_{N}\left( z~q^{s-r+j}\right) %
\right] -q^{-N}\sum_{j=1}^{r}\left[ \left( -1\right) ^{j}a_{j}\left( 
\underline{\alpha }\right) p_{N}\left( z~q^{s-r+j+1}\right) \right] \right\}
=0~. &&  \label{qdifferEq}
\end{eqnarray}%
Next, let us look at this formula---which is a polynomial equation in $z$ of
degree $N+1$---at $z=\zeta _{n},$ where $\zeta _{n}$ is one of the $N$ zeros
of the polynomial $p_{N}\left( z\right) ,$ see (\ref{Zeros}). It is then
plain that there hold the $N$ algebraic equations 
\end{subequations}
\begin{eqnarray}
&&-p_{N}\left( \zeta _{n}~q\right) +\sum_{k=1}^{s}\left( -q\right)
^{-k}b_{k}\left( \underline{\beta }\right) ~\left[ p_{N}\left( \zeta
_{n}~q^{k}\right) -p_{N}\left( \zeta _{n}~q^{k+1}\right) \right]  \notag \\
&&-\left( -1\right) ^{r-s}~\zeta _{n}~\left[ p_{N}\left( \zeta
_{n}~q^{s-r}\right) -q^{-N}p_{N}\left( \zeta _{n}~q^{s-r+1}\right) \right] 
\notag \\
&&-\left( -1\right) ^{r-s}~\zeta _{n}~\left\{ \sum_{j=1,~j\neq r-s}^{r}\left[
\left( -1\right) ^{j}a_{j}\left( \underline{\alpha }\right) ~p_{N}\left(
\zeta _{n}~q^{s-r+j}\right) \right] \right.  \notag \\
&&\left. -q^{-N}\sum_{j=1,~j\neq r-s-1}^{r}\left[ \left( -1\right)
^{j}a_{j}\left( \underline{\alpha }\right) p_{N}\left( \zeta
_{n}~q^{s-r+j+1}\right) \right] \right\} =0~,  \notag \\
&&n=1,2,..,N~.
\end{eqnarray}

And it is then immediately seen that the insertion in these formulas of the
expression (\ref{Zerosb}) of the \textit{monic} polynomial $p_{N}\left(
z\right) $ in terms of its zeros yields (\ref{Prop1}). \textbf{Proposition
2.1} is thereby proven.

\end{document}